\documentclass[twocolumn,amsmath,amssymb,superscriptaddress,prl,floatfix]{revtex4-1}
\usepackage[]{graphicx}
\usepackage[usenames,dvipsnames]{color}
\usepackage{soul}
\usepackage{bm}

\usepackage{times}
\usepackage{amsfonts}
\usepackage{mathrsfs}
\usepackage{graphicx}
\usepackage{dcolumn}
\usepackage{bm}
\usepackage{color}

\usepackage[colorlinks,bookmarks=false,citecolor=blue,linkcolor=red,urlcolor=blue]{hyperref}
\usepackage{multirow}

\begin{document}

\title{Surface $s$-wave superconductivity for oxide-terminated infinite-layer nickelates}

\author{Xianxin Wu}
\email{xianxinwu@gmail.com}
\affiliation{Max-Planck-Institut f\"{u}r Festk\"{o}rperforschung, Heisenbergstrasse 1, D-70569 Stuttgart, Germany}
\affiliation{Beijing National Laboratory for Condensed Matter Physics,
and Institute of Physics, Chinese Academy of Sciences, Beijing 100190, China}

\author{Kun Jiang}
\affiliation{Beijing National Laboratory for Condensed Matter Physics,
and Institute of Physics, Chinese Academy of Sciences, Beijing 100190, China}

\author{Domenico Di Sante}
\affiliation{Institut f\"{u}r Theoretische Physik und Astrophysik and W\"{u}rzburg-Dresden Cluster of Excellence ct.qmat, Universit\"{a}t W\"{u}rzburg, Am Hubland Campus S\"{u}d, W\"{u}rzburg 97074, Germany}

\author{Werner Hanke}
\affiliation{Institut f\"{u}r Theoretische Physik und Astrophysik and W\"{u}rzburg-Dresden Cluster of Excellence ct.qmat, Universit\"{a}t W\"{u}rzburg, Am Hubland Campus S\"{u}d, W\"{u}rzburg 97074, Germany}

\author{A. P. Schnyder}
\affiliation{Max-Planck-Institut f\"{u}r Festk\"{o}rperforschung, Heisenbergstrasse 1, D-70569 Stuttgart, Germany}

\author{Jiangping Hu}
\affiliation{Beijing National Laboratory for Condensed Matter Physics,
and Institute of Physics, Chinese Academy of Sciences, Beijing 100190, China}
\affiliation{CAS Center of Excellence in Topological Quantum Computation and Kavli Institute of Theoretical Sciences,
	University of Chinese Academy of Sciences, Beijing 100190, China}

\author{Ronny Thomale}
\affiliation{Institut f\"{u}r Theoretische Physik und Astrophysik and W\"{u}rzburg-Dresden Cluster of Excellence ct.qmat, Universit\"{a}t W\"{u}rzburg, Am Hubland Campus S\"{u}d, W\"{u}rzburg 97074, Germany}

\date{\today}

\begin{abstract}
We analyze the electronic structure of different surface terminations for infinite-layer nickelates. Surface NiO$_2$ layers are found to be buckled, in contrast to planar bulk layers. While the rare-earth terminated surface fermiology is similar to the bulk limit of the nickelates, the NiO$_2$ terminated surface band structure is significantly altered, originating from the effect of absence of rare-earth atoms on the crystal field splitting. Contrary to the bulk Fermi surfaces, there are two Ni-$3d$ Fermi pockets, giving rise to enhanced spectral weight around the $\bar{\text{M}}$ point in the surface Brillouin zone. From a strong-coupling analysis, we obtain dominant extended $s$-wave superconductivity for the surface layer, as opposed to $d$-wave for the bulk. This finding distinguishes the nickelates from isostructural cuprates, where the analogous surface pairing mechanism is less pronounced. Our results are consistent with region-dependent gap structures revealed in recent STM measurements and provide an ansatz to interpret experimental data of surface-sensitive measurements on the infinite-layer nickelates.
\end{abstract}

\maketitle

{\it Introduction --} The recent discovery of superconductivity in Sr-doped infinite-layer rare-earth nickelates  (Nd,Pr)NiO$_2$ ~\cite{Li2019,pr-hwang} has expanded the range of material classes that are promising hosts for unconventional superconductivity. Starting from the cuprates for which overwhelming experimental evidence is consistent with a $d$-wave order parameter~\cite{TSUEI20001625}, iron pnictides are assumed to predominantly exhibit unconventional $s$-wave order. There, the $s$-wave superconducting order is microscopically preferred by the multi-pocket fermiology~\cite{RevModPhys.83.1589,Hirschfeld2011}. Even though the field is still too young to draw concise connections to more established unconventional superconductors, the infinite-layer nickelates seemingly present themselves as a hybrid between cuprates and pnictides: while the NiO$_2$ layers share close structural similarity to the cuprates, their multi-orbital nature~\cite{Goodge2020} appears more related to the iron pnictides.

 The band structure of RNiO$_2$ (R=rare earth) crystals is of 3D dispersive type with a dominant Ni $d_{x^2-y^2}$ hole band and two R $5d$ electron bands~\cite{Lee2004,Botana,Hepting2020,sakakibara2019model,WuPRB2020} or a R $d_{z^2}$ electron band and an interstitial $s$ electron band~\cite{Nomura2019,Gao2019,Guyh2020}. The theoretical understanding of the correlated electronic structures and pairing states of the nickelates remains controversial so far. Because of the large charge-transfer energy between the Ni-$3d$ and O-$2p$ states in RNiO$_2$, the system is suggested to be in the Hubbard regime, distinct from the charge-transfer nature of cuprates~\cite{JiangM2020,Lechermann2020,Goodge2020}. Recent comprehensive dynamical mean-field theory (DMFT) and GW calculations, however, suggest that both RNiO$_2$ and CaCuO$_2$ are charge-transfer materials~\cite{Karp2020,Olevano2020,SanteD2020}. Moreover, the multi-orbital nature is genuine in both undoped and
hole-doped NdNiO$_2$~\cite{Goodge2020,Petocchi2020}, and Hund's metal physics is likewise argued to be relevant~\cite{WangY2020,KangB2020}. With respect to pairing symmetry, the bulk phase is expected to show a robust $d$-wave pairing, similar to the cuprates~\cite{sakakibara2019model,WuPRB2020,HuLH2019} in a predominantly electronically mediated mechanism. Within a $K-t-J$ model~\cite{ZhangGM2020}, the pairing state is found to be ($d+is$)-wave in the small doping region and changes to a $d$-wave pairing at larger doping~\cite{WangZ2020}.

Until today, the complicated sample growth has prevented a better experimental grasp of the detailed properties of infinite-layer nickelates. Recent experiments revealed an intriguing double-peak T$_c$ dome and weakly insulating behaviors on both sides of the dome~\cite{li2020superconducting,ZengSW2020}. It is challenging to perform and interpret surface-sensitive spectroscopy on the nickelates, such as stemming from angle-resolved photo emission (ARPES), which has not yet been reported, or scanning tunneling microscopy (STM)~\cite{gu2020superconducting}. These experiments would always be particularly sensitive to the electronic structure near the surface, which was likewise observed in cuprates and iron pnictides~\cite{LvYF2015,ZhongY2016,YinJX2016}. Due to the strong hybridization between rare-earth atoms and NiO$_2$ layers~\cite{Nomura2019,Guyh2020}, the electronic structure of surface NiO$_2$ layers can be dramatically distinct from bulk layers. Moreover, the mechanism behind the recent observation of mixed $s$-wave and $d$-wave gap signatures on a rough surface in STM is still unresolved~\cite{gu2020superconducting}. Therefore, the surface modelling of the infinite-layer nickelates can be helpful for a better understanding of the present and future experimental evidences on electronic structure and superconductivity.

\begin{figure}[tb]
\centerline{\includegraphics[width=1.0\columnwidth]{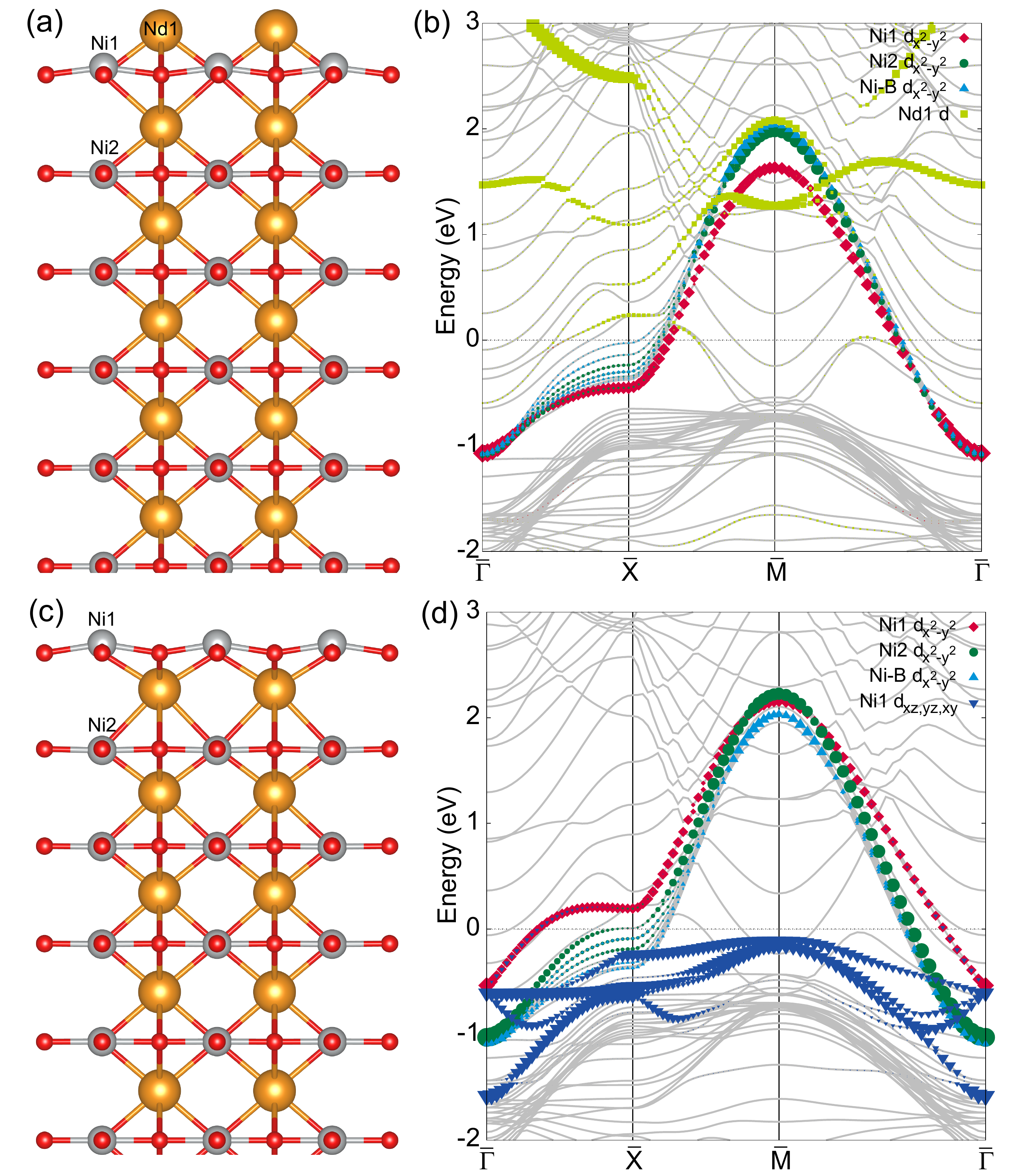}}
\caption{ Crystal structures for (a) the Nd terminated surface and (c) the NiO$_2$ terminated surface of NdNiO$_2$. (b) and (d) are the corresponding orbital resolved band structures. "Ni-B" denotes the bulk Ni atom.
 \label{band} }
\end{figure}

In this Letter, we analyse the electronic structure of different surface terminations for infinite-layer nickelates. From the given layered structure and assuming the clean material limit, two different terminations with either a final rare earth layer or NiO$_2$ layer have to be investigated. From an {\it ab initio} treatment of the two different structures, we obtain effective surface fermiologies for both terminations, which we then further investigate with respect to the onset of superconductivity. The rare-earth termination is dominated by a single Ni-type $d_{x^2-y^2}$ Fermi pocket with slight electron doping [Fig.\ref{band} (b)]. This makes the surface band structure appear similar to the dominant bulk bands, suggesting the onset of $d$-wave superconductivity. The NiO$_2$ termination, however, effectively appears significantly more hole doped than the bulk bands [Fig.\ref{band} (d)]. Owing to the modified crystal-field splitting effect due to absence of the upper rare-earth atoms, a second Ni-type Fermi pocket appears around the $\bar{\text{M}}$ point of the surface Brillouin zone, suggesting a multi-orbital nature. Based on a mean-field $t$-$J$ model analysis we find that this two-pocket fermiology naturally favors an extended $s$-wave superconducting order parameter. It yields a full gap along with a sign change between the two Fermi pockets. We further show that the strong effect arising from the surface termination is specific to the infinite-layer nickelate superconductors, while the isostructural cuprates do not exhibit such a strong propensity. Moreover, we analyze the scenario of Josephson coupling between an $s$-wave surface and a $d$-wave sub-surface state or rough surface islands of different termination, which in the clean limit could yield ($s+id$)-wave pairing in order to maximize the condensation energy within the two layers. In the light of these findings, we give an interpretation of recent STM measurements on rough NdNiO$_2$ surfaces with admixed regions of rare earth and NiO$_2$ terminations. Furthermore, we outline experiments, such as, reflectivity measurements to investigate possible Josephson plasma modes or Kerr spectroscopy to detect time-reversal symmetry breaking without parity breaking in a disordered $s+id$ state.

{\it Electronic structures of the two surface terminations --}
For NdNiO$_2$, there are two kinds of surface terminations, namely, Nd terminated and NiO$_2$ terminated surfaces, as shown in Fig. \ref{band}(a) and \ref{band}(c), respectively. The electronic structure of the NdNiO$_2$/SrTiO$_3$ interface have been theoretically studied~\cite{Bernardini2020,Geisler2020,HeR2020}. In our calculations we modeled these surfaces using slabs of eight NiO$_2$ layers and eight or nine Nd layers plus a vacuum layer of 15 \AA~. In the relaxation, the bottom five layers are fixed while the others are allowed to move freely. The details of calculations can be found in Sec. I of the supplementary materials (SM). After relaxation, one prominent feature is that the surface NiO$_2$ layers are not planar but buckled in both terminations, distinct from bulk layers, which is in response to the internal electric field build-up in the polar slabs~\cite{Geisler2020,HeR2020}.

For the Nd terminated surface, the Ni height with respect to the oxygen plane is smaller of about 0.25 \AA~, owing to the hybridization between top Nd and oxygen atoms. The orbitally resolved band structure of Fig.\ref{band} (b) shows that the Ni $d_{x^2-y^2}$ band of the top NiO$_2$ layer is slightly electron-doped, resembling the bulk band in the $k_z=0$ plane. The bands in gray lines crossing the Fermi level around $\bar{\Gamma}$ and $\bar{\text{M}}$ points are attributed to bulk Nd $d$ oribitals~\cite{Lee2004,Botana,Hepting2020,sakakibara2019model,WuPRB2020}. Similar to bulk layers, other Ni $d$ orbitals are located far below the Fermi level and surface Nd atoms have some contributions to the Fermi surfaces.

For the NiO$_2$ terminated surface, the Ni height with respect to the oxygen plane is larger and about 0.30 \AA~ in the surface layer. The underlying Nd layer further moves closer to the top NiO$_2$ layer with a vertical separation of 1.23 \AA~ and couples strongly with oxygen atoms (Fig.\ref{band}(c)). The corresponding orbitally resolved band structure is displayed in Fig.\ref{band}(d). We find that the $d_{x^2-y^2}$ band of the surface NiO$_2$ (red diamonds) is heavily hole-doped compared with bulk NiO$_2$ layers (blue up-triangles) due to the absence of the charge reservoir layer, i.e., rare-earth atoms. Moreover, it also significantly affects the crystal field splitting and pushes upwards Ni $d_{xz/yz}$
and $d_{xy}$ orbitals. The onsite energies of the $d_{xz/yz}$ and the $d_{xy}$ orbitals relative to the $d_{x^2-y^2}$ orbital on the surface layer increase by 0.45 and 0.16 eV, respectively, compared with bulk layers. Therefore, the Ni $d_{xy}$, $d_{xz/yz}$ orbitals have strong hybridizations with $d_{x^2-y^2}$ orbitals near the Fermi level and their bands are located just slightly below the Fermi level, especially around the $\bar{\text{M}}$ point. Actually, these bands cross the Fermi level without surface atomic relaxation (see Sec. I of SM). Without external doping, there is only a small electron pocket around the $\bar{\Gamma}$ point for the surface NiO$_2$ layer. In experiments, superconductivity was achieved by hole doping, introduced by the substitution of Nd by Sr\cite{Li2019,pr-hwang}. Therefore, the multi-orbital nature is expected to be relevant for the hole-doped NiO$_2$ surface; this is distinct from bulk NiO$_2$ layers, where only Ni $d_{x^2-y^2}$ orbital plays an essential role~\cite{sakakibara2019model,WuPRB2020}.

As the rare-earth atoms are not ionic charge carriers and couple strongly to NiO$_2$ layers in infinite-layer RNiO$_2$, the multi-orbital nature is a unique feature on its NiO$_2$-terminated surface. To demonstrate this, we also performed similar calculations for their cuprate analog CaCuO$_2$ whose band structures for the two terminations are provided in the Sec. II of SM. There, for both terminations, only the Cu $d_{x^2-y^2}$ orbital of the surface CuO$_2$ layer dominates near the Fermi level, while other Cu $d$ orbitals are well below the Fermi level in accordance with its charge-transfer nature.

\begin{figure}[tb]
\centerline{\includegraphics[width=1.0\columnwidth]{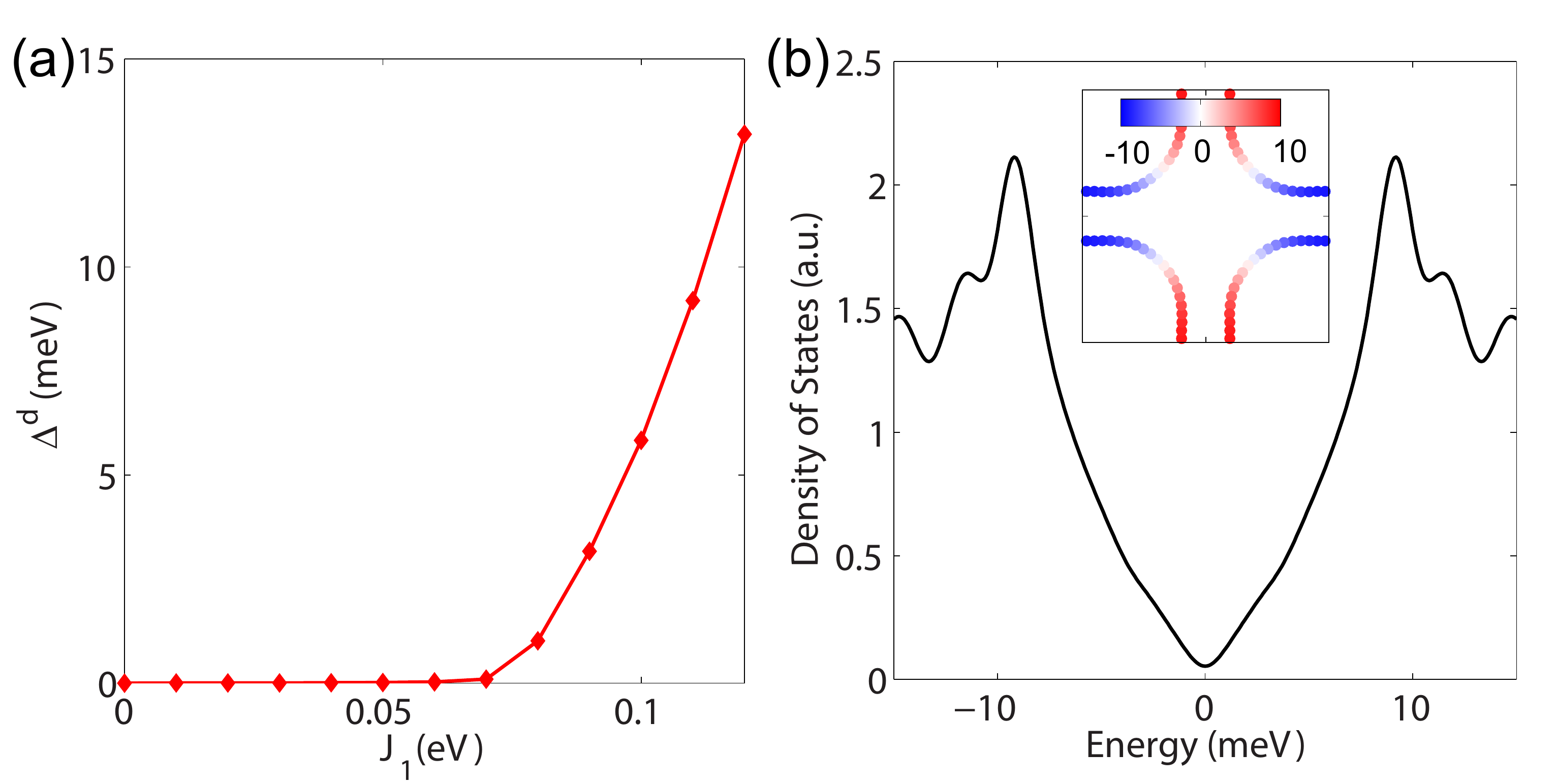}}
\caption{(a) The absolute value of $d$-wave gap as a function of the exchange coupling parameter $J_1$ for the Nd terminated surface of Nd$_{0.8}$Sr$_{0.2}$NiO$_2$. (b) Density of states of the $d$-wave pairing state at the Nd terminated surface for $J_1 =0.1$ eV. The inset shows the momentum dependence of the gap along the Fermi surface in units of meV.
 \label{surfNd} }
\end{figure}

{\it Pairing states for surface layers --} For the bulk system, a robust $d_{x^2-y^2}$-wave pairing is found for Nd$_{1-x}$Sr$_x$NiO$_2$, dominantly contributed by the Ni $d_{x^2-y^2}$-orbital Fermi surface~\cite{sakakibara2019model,WuPRB2020}. The surface layers can favor distinct pairing states from the bulk, owing to their different carrier dopings and electronic structures. For the Nd terminated surface, despite slight electron doping, the Fermiology of surface layer is close to that of a bulk layer for the $k_z=0$ plane. Therefore, the surface layer is expected to favor a $d_{x^2-y^2}$ pairing as well. We perform numerical calculations to verify it: we first construct the one band tight-binding (TB) model (Sec. III of SM) to describe the electronic structure of the surface NiO$_2$ layer and then study the pairing within a similar t-J model as Ref.~\cite{WuPRB2020}. Fig.\ref{surfNd}(a) displays the $d$-wave gap as a function of the exchange parameter and the $s$-wave gap vanishes, indicating that the $d$-wave pairing is dominant. The corresponding density of states (DOS), shown in Fig.\ref{surfNd}(b), exhibits a V-shaped gap structure, similar to bulk NiO$_2$ layers or typical cuprates.

\begin{figure}[tb]
\centerline{\includegraphics[width=1.0\columnwidth]{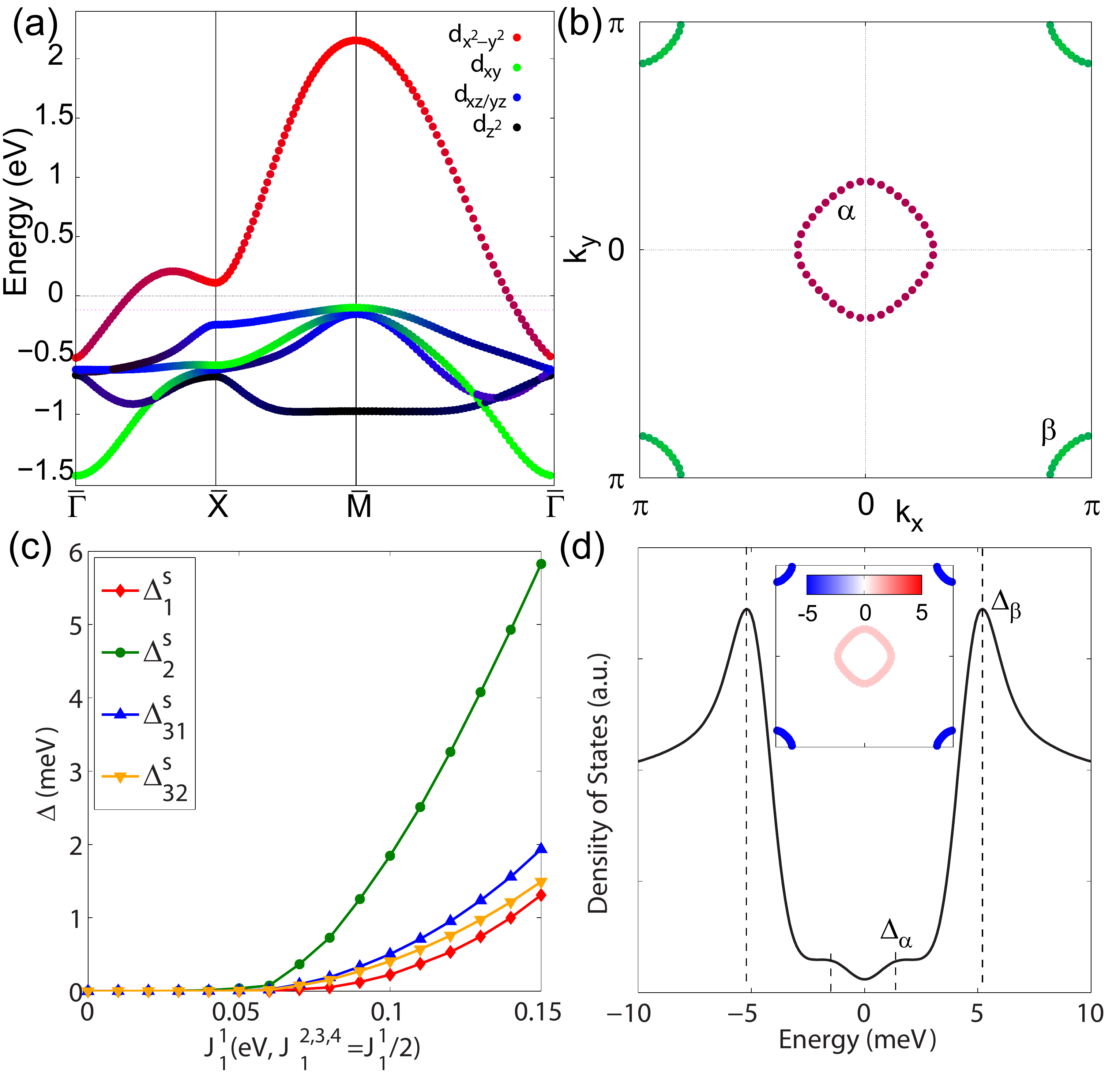}}
\caption{(a) Orbitally resolved band structure of the TB model for the surface NiO$_2$ layer. The pink dashed line denotes the chemical potential for hole doping $\delta=0.12$. (b) Representative Fermi surfaces with orbital characters for the hole-doped surface NiO$_2$ layer of Nd$_{0.8}$Sr$_{0.2}$NiO$_2$ (hole doping level is $\delta=0.12$). (c) Orbital dependent superconducting gaps as a function of exchange coupling parameters for $\delta=0.12$ with $J^{2,3,4}_1=J^1_1/2$. (d) Density of states of the extended $s$-wave pairing state in the NiO$_2$ surface layer for $\delta$ = 0.12 and $J_1^1$ = 0.12 eV. The inset shows the momentum dependance of the gap along the FSs in units of meV.
 \label{caltj} }
\end{figure}

For the NiO$_2$ terminated surface, however, other $d$ orbitals than $d_{x^2-y^2}$ can likewise contribute to the fermiology of the top NiO$_2$ layer, which may result in a change of pairing symmetry. We first study the multi-orbital TB model for the distorted NiO$_2$ layers with heavy hole doping and introduce the operator $\psi^\dag_{\textbf{k}\sigma}=[c^\dag_{1\sigma}(\textbf{k}),c^\dag_{2\sigma}(\textbf{k}),c^\dag_{3\sigma}(\textbf{k}),c^\dag_{4\sigma}(\textbf{k}),c^\dag_{5\sigma}(\textbf{k})]$, where $c^\dag_{\alpha\sigma}(\textbf{k})$ is a Fermionic creation operator with  $\sigma$ and $\alpha$ denoting spin and orbital indices, respectively. The orbital index $\alpha=1,2,3,4,5$ represents the Ni $d_{x^2-y^2}$ for 1, $d_{xy}$ for 2, $d_{xz}$ for 3, $d_{yz}$ for 4 and $d_{z^2}$ for 5, respectively. The TB Hamiltonian can be written as,
 \begin{eqnarray}
 H_{0}=\sum_{\textbf{k}\sigma}\psi^\dag_{\textbf{k}\sigma}(h(\textbf{k})-\mu)\psi_{\textbf{k}\sigma}.
 \end{eqnarray}
  The matrix elements in the Hamiltonian $h(\textbf{k})$ matrix are given in the Sec. III of SM. The orbitally resolved band structure from the above TB model, showing good agreement with DFT calculations (Sec. III of SM), is displayed in Fig.\ref{caltj}(a). In experiments, a superconducting dome was found in Nd$_{1-x}$Sr$_x$NiO$_2$ for $0.12\leq x\leq 0.25$~\cite{li2020superconducting}. Assuming that charge transfer from (Nd,Sr) layers to NiO$_2$ layers is inversely proportional to their separations, the hole doping of surface NiO$_2$ layer is estimated to be $0.08\leq \delta \leq 0.16$. Within this doping level, there are an electron pocket around the $\bar{\Gamma}$ point and a hole pocket around the $\bar{\text{M}}$ point, attributed to Ni $d_{x^2-y^2}$, $d_{xy}$ and $d_{xz/yz}$ orbitals. These Fermi surfaces can also be reproduced in Nd$_{0.8}$Sr$_{0.2}$NiO$_2$ in first-principles calculations (Sec. I of SM). Since the bands around the $\bar{\text{M}}$ point are flat, the corresponding hole pocket has a large DOS. Representative Fermi surfaces with a hole doping $\delta=0.12$ (for Nd$_{0.8}$Sr$_{0.2}$NiO$_2$) are shown in Fig.\ref{caltj}(b), suggesting that the multi-orbital nature is one prominent feature for surface NiO$_2$ layers.

   Due to the appearance of the hole pocket around $\bar{\text{M}}$ with an enhanced spectral weight, distinct from over-doped cuprates, and in order to account for the multi-orbital nature of the surface NiO$_2$ layers, we adopt a multi-orbital t-J model to study pairing propensities. Similar to iron-based superconductors~\cite{SiQM2008,Seo2008}, we consider for the surface NiO$_2$ layers inplane antiferromagnetic coupling between the Ni spins,
\begin{eqnarray}
H_{J}=\sum_{\langle ij\rangle\alpha}J^\alpha_{ij}(\mathbf{S}_{i\alpha}\mathbf{S}_{j\alpha}-\frac{1}{4}n_{i\alpha}n_{j\alpha})
\end{eqnarray}
where
$\bm{S}_{i\alpha}=\frac{1}{2}c_{i\alpha\sigma}^{\dagger}\bm{\sigma}_{\sigma\sigma'}c_{i\alpha\sigma'}$
is the local spin operator and $n_{i\alpha}$is the local density
operator  for Ni $\alpha$ orbital ($\alpha=x^2-y^2,xy,xz,yz,z^2$). $\langle ij\rangle$ denotes the in-plane nearest neighbors and the in-plane coupling is $J^\alpha_x=J^\alpha_y=J^\alpha_1$. By performing the Fourier transformation, $H_{J}$ in momentum space reads
\begin{eqnarray}
H_{J}=\sum_{\alpha\mathbf{k,k}'}V^\alpha_{\mathbf{k},\mathbf{k}'}c_{\bm{k}\alpha\uparrow}^{\dagger}c_{-\bm{k}\alpha\downarrow}^{\dagger}c_{-\bm{k}'\alpha\downarrow}
c_{\bm{k}'\alpha\uparrow},
\end{eqnarray}
with $V^\alpha_{\mathbf{k},\mathbf{k}'}=-\frac{2J^\alpha_1}{N}\sum_{\pm}(cosk_x\pm cosk_y)(cosk'_x\pm cosk'_y)$ and $N$ being the number of lattice sites. Here we investigate the pairing state for the doped system and neglect the
no-double-occupancy constraint on this t-J model and then perform a mean-field decoupling (details can be found in Sec. IV of SM). In the calculations, we only consider the intraorbital pairing and neglect inter-orbital pairing, as the orbital mixture on Fermi surfaces is not strong. The orbital dependent gap is defined as $ \Delta^{s/d}_{\alpha} = -\frac{2J_1}{N}\sum_{\mathbf{k}'}\langle
c_{-\mathbf{k}'\alpha\downarrow}c_{\mathbf{k}'\alpha\uparrow}\rangle (cosk'_x\pm cosk'_y)$ and we further introduce $\Delta^{s/d}_{31}=\frac{1}{2}(\Delta^{s}_{3}\pm \Delta^{s}_{4}+\Delta^{d}_{3}\mp \Delta^{d}_{4}) $ and $\Delta^{s/d}_{32}=\frac{1}{2}(\Delta^{s}_{3}\pm \Delta^{s}_{4}-\Delta^{d}_{3}\pm \Delta^{d}_{4}) $ for $d_{xz/yz}$ orbitals.
The obtained gaps as a function of exchange coupling parameters at $\delta=0.12$ are shown in Fig.\ref{caltj}(c) and the doping dependent gaps are given in Sec. IV of SM. We find that the extended $s$-wave pairing is always dominant within the estimated hole doping region, and that the gap of $d_{xy}$ orbital is much larger than other orbitals, originating from its enhanced DOS around $\bar{\text{M}}$. Moreover, as displayed in the inset of Fig.\ref{caltj}(d), there is a sign change in the gap function between the electron pocket around $\bar{\Gamma}$ and the hole pocket around $\bar{\text{M}}$, in analogy to $s_{\pm}$ pairing in iron based superconductors~\cite{Hirschfeld2011} and overdoped CuO$_2$ monolayers~\cite{JiangK2018}. The corresponding DOS in Fig.\ref{caltj}(d) exhibits a general U-shaped structure and two-gap feature.

\begin{figure}[tb]
\centerline{\includegraphics[width=1.0\columnwidth]{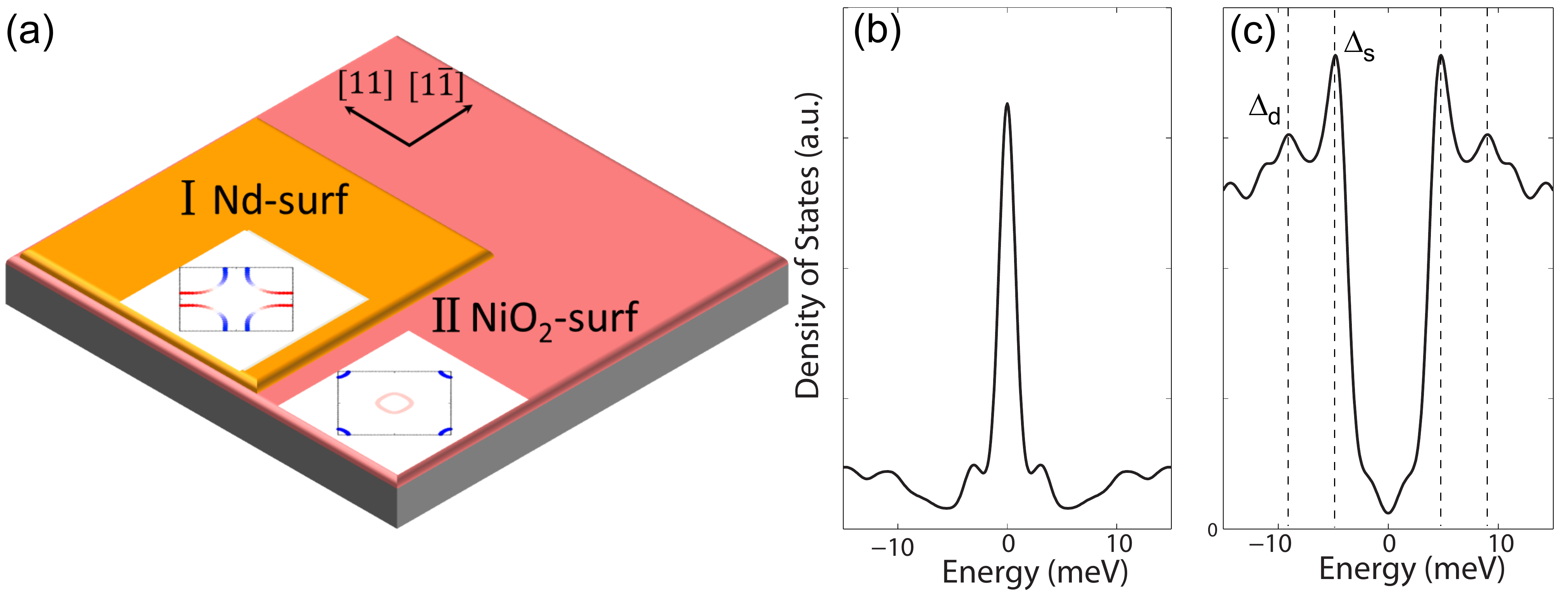}}
\caption{(a) Schematic for the different surface terminations of NdNiO$_2$. (b) DOS at the domain boundary between Nd-terminated and NiO$_2$-terminated surfaces along the [$11$] or [$1\bar{1}$] direction. (c) DOS for the $s_{\pm}-id_{x^2-y^2}$ pairing state including the coupling between surface and bulk NiO$_2$ layers. \label{dossid} }
\end{figure}

{\it Termination-dependent gap structures --} In experiments, two terminations are expected to be randomly distributed on the surface. In the (Sr,Nd) terminated surface, a V-shaped gap is expected in the tunneling spectrum of STM measurements. For the NiO$_2$ terminated surface, however, the surface NiO$_2$ layer has $s_{\pm}$-wave pairing, inducing a U-shaped gap. In the domain wall between two kinds of terminations, shown in Fig.\ref{dossid}(a), zero-energy Andreev flat bands will appear in the [$11$] or [$1\bar{1}$] direction, inducing a sharp zero-bias peak in the local DOS of Fig.\ref{dossid}(b). Note that the extensive ground-state degeneracy of the Andreev flat bands is thermodynamically unstable. I.e., at low enough temperatures it will be lifted by residual interactions, which induce a symmetry-broken state at the edge, e.g., ($d+is$)-wave pairing at the edge, or possibly edge magnetism \cite{Honerkamp2000,Hofmann2016}. This should be observable in experiments in terms of a splitting of the zero-bias peak at low temperatures.

As the surface NiO$_2$ layer has $s_{\pm}$-wave pairing and the bulk NiO$_2$ layer favors a $d_{x^2-y^2}$-wave pairing, the coupling between them introduces a Josephson junction along the $c$ axis. While this scenario is not what is given in current experiments, one might reach it in future samples due to the enhancement of synthesis quality. Generally, the free energy of a Josephson junction~\cite{Golubov2004,Goldobin2007,YangZS2018} can be written as,
\begin{eqnarray}
F=F_0-g_1cos\Delta \phi+g_2cos2\Delta \phi,
\end{eqnarray}
where $\Delta \phi$ is the phase difference between the two superconducting order parameters. The first term is independent of the phase difference and the second and third terms are the first order and second order Josephson couplings, respectively. Usually, the conventional Josephson coupling $g_1$ is much larger than $g_2$. However, in our case, as the two superconducting orders belong to two different irreducible representations, $g_1$ should vanish. Therefore, the remaining second order Josephson coupling $g_2$ will lead to $\Delta \phi=\pm \frac{\pi}{2}$ by minimizing the free energy, forming an $s_{\pm}\pm id_{x^2-y^2}$ pairing state depending on the sign of $g_2$. We perform calculations with a t-J model by including both Hamiltonians of surface and bulk NiO$_2$ layers and their hybridizations (details in Sec. IV of SM), and find that the time-reversal-symmetry breaking $s_{\pm}-id_{x^2-y^2}$ pairing state is energetically more favorable. The corresponding DOS is shown in Fig.\ref{dossid}(c), where the gap is generally V-shaped enriched with multiple-gap features. In such a state, supercurrent can be induced near a nonmagnetic impurity and the corners of a square sample, which may be detected by using SQUID~\cite{Lee2009}.

Recent STM measurements seem to reveal an $s$-wave gap and a $d$-wave gap depending on the probed region and a mixture of them in some cases~\cite{GuQQ2020}. In contrast to the explanation given in Ref.~\cite{GuQQ2020}, we do not believe that this observation can be explained by orbitally selective pairing, where the Ni $d_{x^2-y^2}$-derived Fermi surface has a $d$-wave superconducting gap while Nd $d_{xy,z^2}$-derived Fermi surfaces have an $s$-wave superconducting gap~\cite{GuQQ2020}. As the hybridization between Ni $d_{x^2-y^2}$ and Nd $d_{xy,z^2}$ orbitals is strong, these Fermi surfaces will naturally have superconducting gaps with the same symmetry, but different gap sizes. Instead the scenario of different terminations presents on a rough surface seems consistent with the experimental measurements: following our explanation there should be a $d$-wave gap for the (Nd,Sr)-terminated surface and an $s$-wave gap for the NiO$_2$-terminated surface; further including the coupling between surface and bulk NiO$_2$ layers, a mixture of those two superconducting order appears natural to occur for a partially NiO$_2$-terminated surface.

{\it Conclusion --} We investigated the electronic structures for different surface terminations in infinite-layer nickelates and find that the surface NiO$_2$ layer has a buckled structure and differs from the bulk layer in terms of charge doping. The rare earth terminated surface Fermiology is similar to the bulk albeit with a moderate electron doping. For the NiO$_2$ terminated surface, however, the absence of rare-earth atoms significantly affects the crystal field splitting and pushes upwards the Ni $d_{xz/yz}$ and $d_{xy}$ orbitals. Consequently, the surface layer is not only heavily hole-doped but also develops an additional electron pocket around the $\bar{\text{M}}$ point besides the hole pocket around $\bar{\Gamma}$, derived from  $d_{xy}$ and $d_{xz/yz}$ orbitals. From a strong coupling analysis, we find that this two-pocket Fermiology naturally favors
 an $s_{\pm}$-wave pairing, in analogy to iron based superconductors.  Further including the bulk $d_{x^2-y^2}$-wave pairing, the second order Josephson coupling induces a $s_{\pm}- id_{x^2-y^2}$ pairing state in the system. Our results seem to be in line with recent experimental observations of both $s$-wave gap and $d$-wave gap signatures. Our finding provide a starting point for interpreting surface-sensitive spectroscopy data in future experiments on infinite-layer nickelates, such as ARPES or optical spectroscopy, including Kerr measurements to possibly track down the time-reversal symmetry breaking from any kind of surface $s+id$ pairing.

{\it Acknowledgments--} We thank Yujie Sun for helpful discussions. This work is funded by the Deutsche Forschungsgemeinschaft (DFG, German Research Foundation) through Project-ID 258499086 - SFB 1170 and through the W\"{u}rzburg-Dresden Cluster of Excellence on Complexity and Topology in Quantum Matter-ct.qmat Project-ID 390858490 - EXC 2147. J.P.Hu was supported by the Ministry of Science and Technology of China 973 program
(No. 2017YFA0303100), National
Science Foundation of China (Grant No. NSFC11888101), and the
Strategic Priority Research Program of CAS (Grant No.XDB28000000).

{\it Note added: }  We recently became aware of an independent work about nickelate thin films grown on SrTiO$_3$ substrate~\cite{ZhangY2020}, where the surface distortions and the band structure of NiO$_2$ terminated surface are consistent with our paper.

\bibliography{biblio}

\clearpage

\newcommand{\beginsupplement}{%
        \setcounter{table}{0}
        \renewcommand{\thetable}{S\arabic{table}}%
        \setcounter{figure}{0}
        \renewcommand{\thefigure}{S\arabic{figure}}%
        \setcounter{equation}{0}
        \renewcommand{\theequation}{S\arabic{equation}}
     }

\renewcommand{\theequation}{S\arabic{equation}}
\renewcommand{\thefigure}{S\arabic{figure}}
\renewcommand{\bibnumfmt}[1]{[S#1]}
\renewcommand{\citenumfont}[1]{S#1}

\section{details of DFT calculations}
\begin{figure}[tb]
\centerline{\includegraphics[width=0.8\columnwidth]{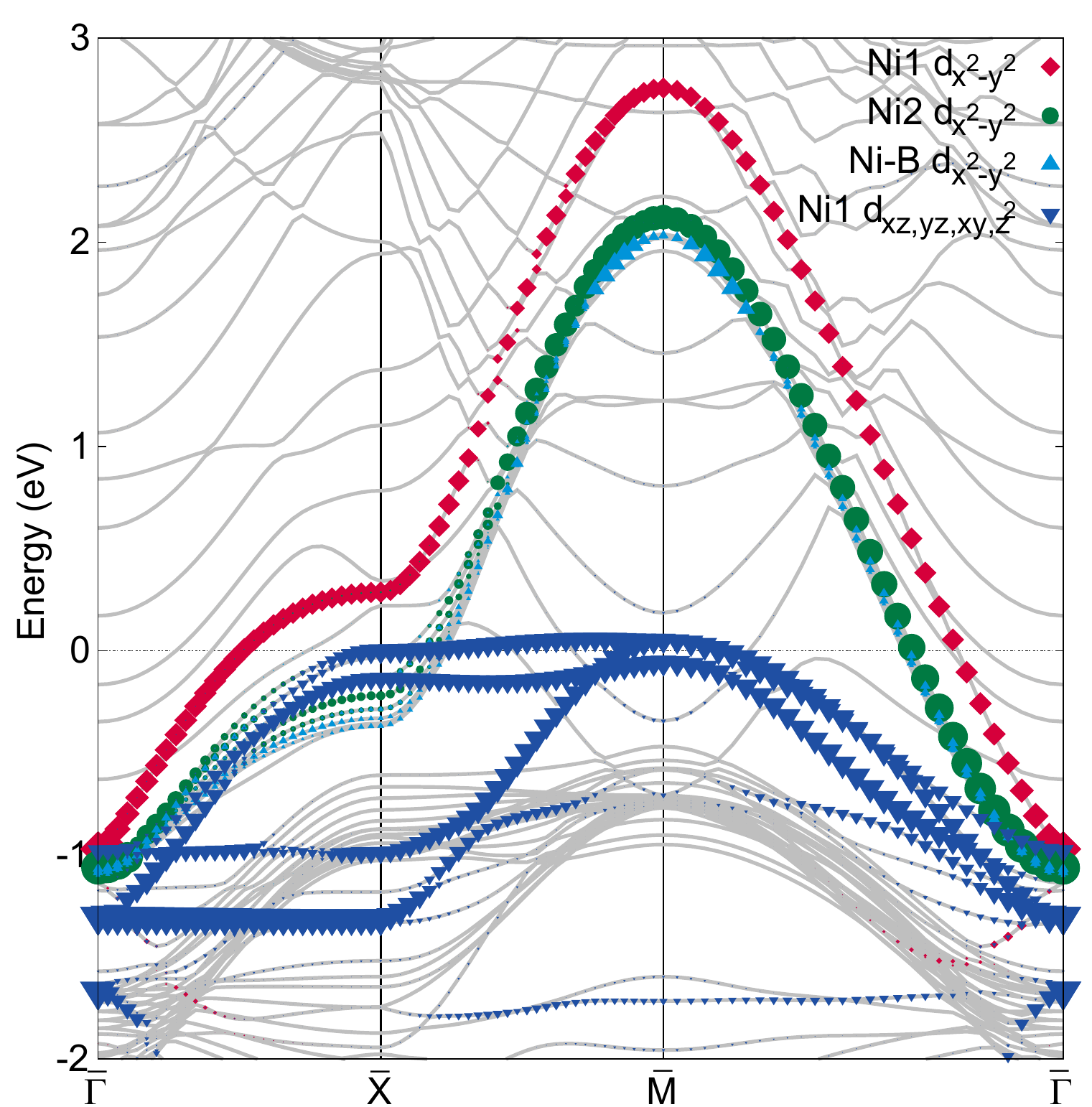}}
\caption{(color online) DFT Band structure for NiO$_2$ terminated surface of NdNiO$_2$ without relaxation.
 \label{dftv1} }
\end{figure}
Our DFT calculations employ the projector augmented
wave (PAW) method encoded in the Vienna {\it ab initio} simulation
package (VASP)\cite{Kresse1993,Kresse1996,Kresse1996-2}, and the generalized-gradient
approximation (GGA)\cite{Perdew1996} for the exchange correlation
functional are used. The cutoff energy of 500 eV is taken
for expanding the wave functions into plane-wave basis and the Nd $4f$ electrons are considered frozen in the core. In calculations for both nickelates and cuprates, we modeled the surfaces using a slab of eight NiO$_2$/CuO$_2$ layers and eight or nine Nd/Ca layers plus a vacuum layer of 15~\AA. The number of layers is large enough to give converged electronic structures of surface layers. The adopted lattice parameters are $a=3.91$~\AA~and $c=3.37$~\AA~for NdNiO$_2$ and $a=3.86$~\AA~and $c=3.20$~\AA~for CaCuO$_2$.
The Brillouin zone is sampled in the $k$ space
within the Monkhorst-Pack scheme\cite{Monkhorst1976}. The number of these
k points is $9\times9\times1$ for two terminations of NdNiO$_2$ and  CaCuO$_2$.  In the relaxation, the bottom five layers
are fixed while the others are allowed to move freely and forces are minimized to less than 0.02 eV/\AA~. We have checked the dipole correction on the atomic relaxation and find that it has negligible effect. After
relaxation, one prominent feature is that surface NiO$_2$/CuO$_2$ layers
are not planar and buckled in both terminations, distinct from
bulk layers.

\begin{figure}[t]
\centerline{\includegraphics[width=1.0\columnwidth]{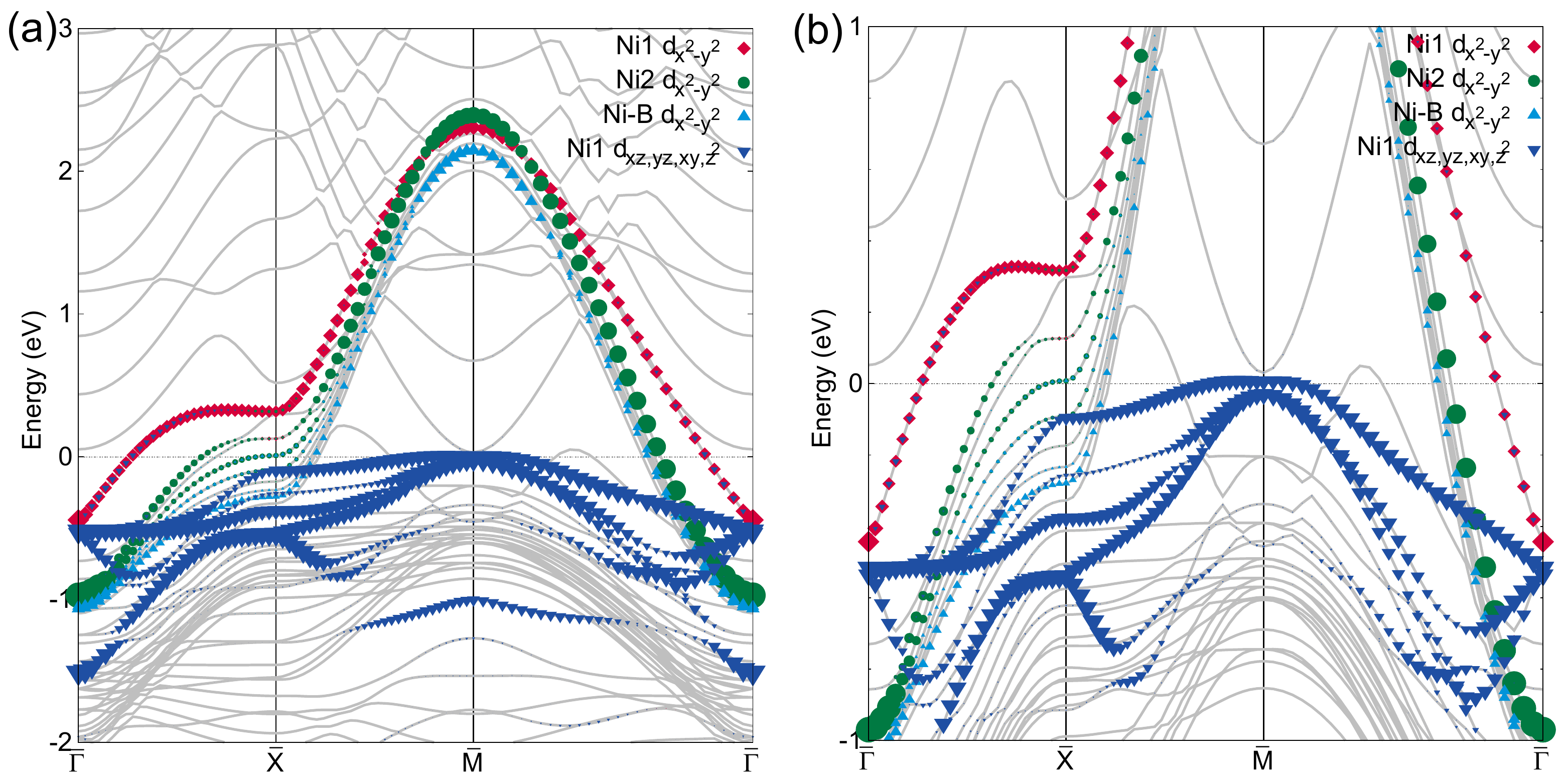}}
\caption{(color online) (a) DFT Band structure for NiO$_2$ terminated surface of Nd$_{0.8}$Sr$_{0.2}$NiO$_2$ with relaxation. (b) A zoom-in band structure. Besides the electron pocket around $\bar{\Gamma}$, there is a hole pocket around $\bar{\text{M}}$.
 \label{dfthole} }
\end{figure}

For NiO$_2$ terminated surface without relaxation, the band structure is shown in Fig.\ref{dftv1}. For the surface NiO$_2$ layer, we find that besides the hole pocket around the $\bar{\Gamma}$ point, there are Fermi surfaces around $\bar{\text{M}}$ point attributed to $d_{xz/yz}$ and $d_{xy}$ orbitals for the surface layer. After relaxation, the lattice distortions push these bands below the Fermi level (see the main text) but Fermi surface around the M point can still appear in hole-doped nickelates (Nd,Sr)NiO$_2$. To demonstrate it, we study Sr doped nickelates Nd$_{0.8}$Sr$_{0.2}$NiO$_2$ using the virtual crystal approximation in DFT calculations and the band structure is displayed in Fig.\ref{dfthole}. There is an additional hole pocket around $\bar{\text{M}}$ point, attributed to $d_{xz/yz}$ and $d_{xy}$ orbitals. This is consistent with the doping using a rigid-band shift of the Fermi level in our tight-binding model in the main text.

\section{band structure for CuO$_2$ and Ca terminated surfaces in CaCuO$_2$}
\begin{figure}[tb]
\centerline{\includegraphics[width=1.0\columnwidth]{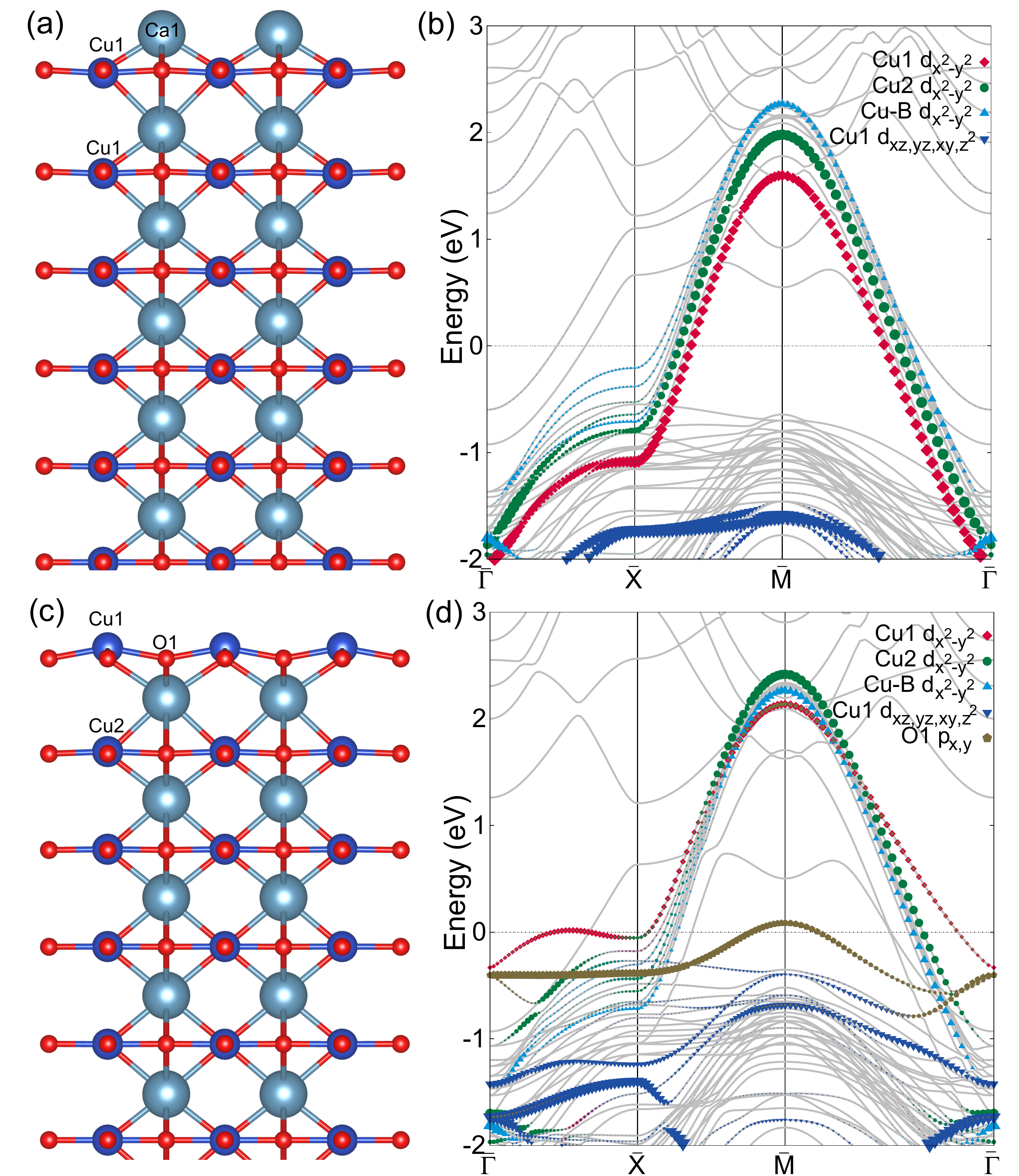}}
\caption{(color online) (a) Crystal structures for Ca terminated (a) and CuO$_2$ terminated (c) surfaces of CaCuO$_2$. The corresponding orbital resolved band structures are in (b) and (d). "Ni-B" denotes the bulk Ni atom.
 \label{bandCu} }
\end{figure}

To demonstrate the unique features of surface NiO$_2$ layers, we also performed calculations for CuO$_2$ and Ca terminated surfaces in a typical infinite-layer cuprate CaCuO$_2$. Similar to nickelates, the surface CuO$_2$ is buckled in both terminations, as shown in Fig.\ref{bandCu} (a) and (c). For Ca terminated surface, the band structure is displayed in Fig.\ref{bandCu} (b) and the surface layer is electron doped, similar with nickelates. From the orbitally resolved band structure of the CuO$_2$ terminated surface in  Fig.\ref{bandCu} (d), the surface layer is heavily hole doped and there is an additional hole pocket attributed to oxygen atoms, indicating that the holes are located at oxygen sites. This is consistent with the charge-transfer nature of cuprates and other Cu $d$ orbitals are well below the Fermi level. These features are in sharp contrast to the case of nickelates, where multi-orbital nature is essential for the surface NiO$_2$ layer. Therefore, the surface pairing propensity in infinite-layer nickelates is quite intriguing.

\section{Tight-binding models of the surface NiO$_2$ layer in Nd terminated and NiO$_2$ terminated surfaces}

\begin{figure}[t]
\centerline{\includegraphics[width=1.0\columnwidth]{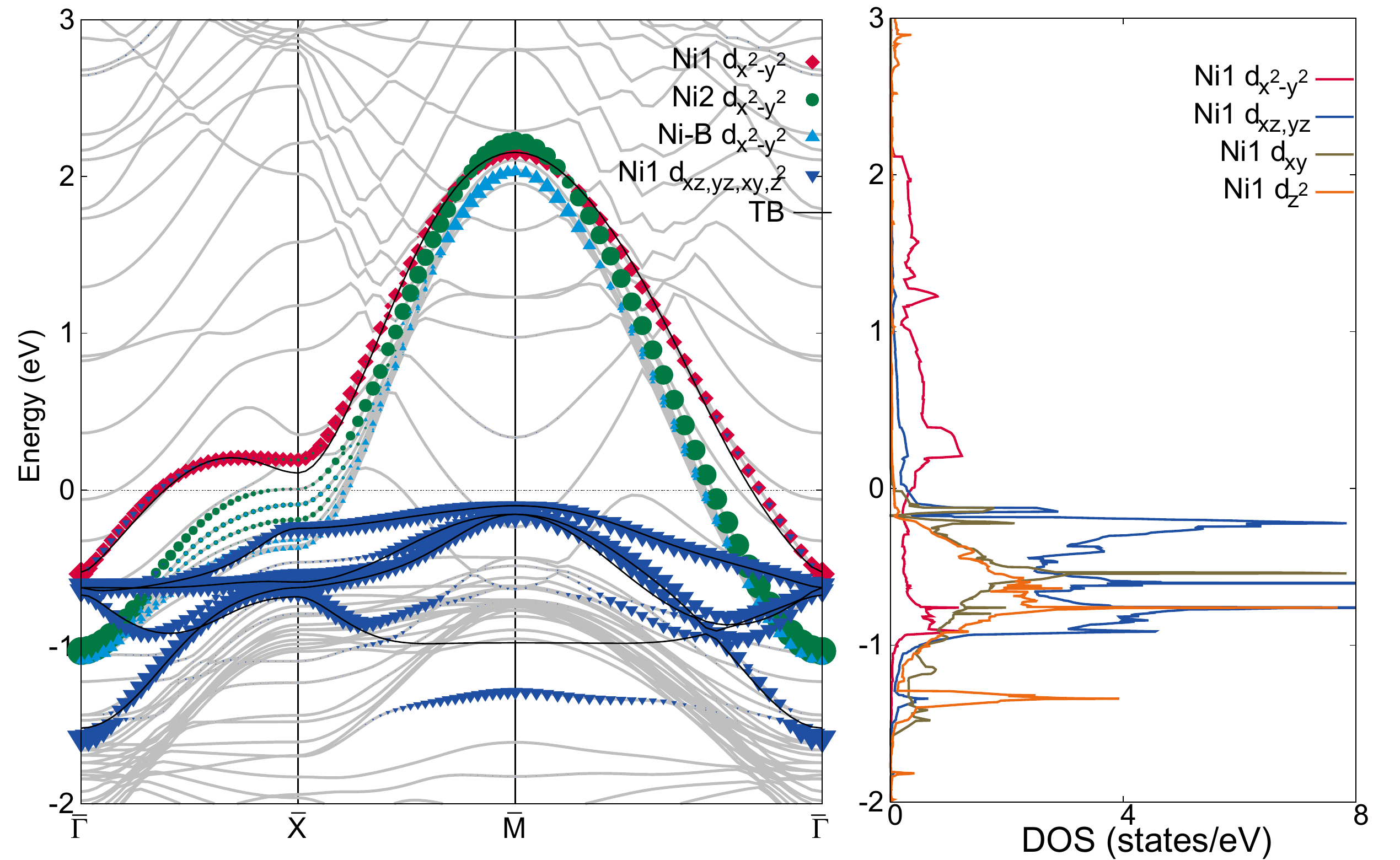}}
\caption{(color online) Band structure from DFT and five-band tight-binding model (blak lines) for NiO$_2$ terminated surface and orbitally resolved DOS for the Ni1 atom.
 \label{dfttb} }
\end{figure}

\begin{table}[t]
\caption{\label{onsite} Calculated on-site energies relative to the $d_{x^2-y^2}$ orbital and hoppings of $d_{x^2-y^2}$ orbitals for surface and bulk NiO$_2$ layer in NdNiO$_2$.  }
\renewcommand{\multirowsetup}{\centering}
\begin{tabular}{ccc}
\hline
\hline
Wannier onsite energies (eV) & bulk layer & surface layer \\
   \hline
 $d_{xz/yz}$  & -1.48 & -1.03    \\
  $d_{xy}$  & -1.38 & -1.22    \\
   $d_{z^2}$  & -1.73 & -1.32    \\
\hline
 hoppings of $d_{x^2-y^2}$ orbital (eV)  &   &  \\
      $t_x$  & -0.3761 & -0.3369   \\
      $t_{xy}$  & 0.0844 & 0.0889    \\
      $t_{xx}$  & -0.0414 & -0.0329    \\

\hline
 \hline
\end{tabular}

\end{table}

For Nd terminated surface, the electronic structure is dominated by the Ni $d_{x^2-y^2}$ orbital and therefore can be described by a one-band tight-binding model. The Hamiltonian matrix can be written as,
\begin{eqnarray}
h_{Nd}(\bm{k})&=&\epsilon_0-\mu+2s^x_{11}(cosk_x+cosk_y)+4s^{xy}_{11}cosk_xcosk_y\nonumber\\
&&+2s^{xx}_{11}(cos2k_x+cos2k_y)\nonumber\\
&&+4s^{xxy}_{11}(cos2k_xcosk_y+cosk_xcos2k_y)\nonumber\\
&&+4s^{xxyy}_{11}cos2k_xcos2k_y,
\end{eqnarray}
with hopping parameters
\begin{eqnarray}
&&s^x_{11}= -0.316575, \quad s^{xy}_{11}=0.092113,\quad s^{xx}_{11}=-0.047588,\nonumber\\
&&s^{xxy}_{11}=-0.004311,\quad s^{xxyy}_{11}=0.003844, \quad \epsilon_0=3.283151,\nonumber\\
\end{eqnarray}
in units of eV. The chemical potential for Nd terminated surfaces of NdNiO$_2$ and  Nd$_{0.8}$Sr$_{0.2}$NiO$_2$ are $\mu=3.1510$ eV and $\mu=2.9710$ eV, respectively.

As discussed in the main text, the surface NiO$_2$ is heavily hole-doped and the corresponding Fermi surfaces are attributed to $d_{xy}$ and $d_{xz/yz}$ orbitals besides $d_{x^2-y^2}$ orbitals, indicating that a multi-orbital TB model is required to capture the electronic structure. The onsite energies of Ni $d$ orbitals in the surface layer is given in Table \ref{onsite}, in comparison to bulk layers. We find that the relative onsite energies of $d$ orbitals relative to $d_{x^2-y^2}$ orbital increase, originating from the absence of rare-earth atoms on the surface. We introduce the operator $\psi^\dag_{\textbf{k}\sigma}=[c^\dag_{1\sigma}(\textbf{k}),c^\dag_{2\sigma}(\textbf{k}),c^\dag_{3\sigma}(\textbf{k}),c^\dag_{4\sigma}(\textbf{k}),c^\dag_{5\sigma}(\textbf{k})]$, where $c^\dag_{\alpha\sigma}(\textbf{k})$ is a Fermionic creation operator with  $\sigma$ and $\alpha$ being spin and orbital indices, respectively. The orbital index $\alpha=1,2,3,4,5$ represent the Ni $d_{x^2-y^2}$ for 1, $d_{xy}$ for 2, $d_{xz}$ for 3, $d_{yz}$ for 4 and $d_{z^2}$ for 5, respectively. The tight-binding Hamiltonian can be written as,

 \begin{eqnarray}
 H_{0}=\sum_{\textbf{k}\sigma}\psi^\dag_{\textbf{k}\sigma}(h(\textbf{k})-\mu)\psi_{\textbf{k}\sigma}.
 \end{eqnarray}
  The matrix elements in the Hamiltonian $h(\textbf{k})$ matrix are given by,
  \begin{widetext}
\begin{eqnarray}
\label{caas_tb}
h_{11}(\bm{k})&=&\epsilon_1+2t^x_{11}(cosk_x+cosk_y)+4t^{xy}_{11}cosk_xcosk_y+2t^{xx}_{11}(cos2k_x+cos2k_y)+4t^{xxy}_{11}(cos2k_xcosk_y+cosk_xcos2k_y)\nonumber\\
&&+4t^{xxyy}_{11}cos2k_xcos2k_y\\
h_{22}(\bm{k})&=&\epsilon_2+2t^x_{22}(cosk_x+cosk_y)+4t^{xy}_{22}cosk_xcosk_y+2t^{xx}_{22}(cos2k_x+cos2k_y)+4t^{xxy}_{22}(cos2k_xcosk_y+cosk_xcos2k_y)\nonumber\\
&&+4t^{xxyy}_{22}cos2k_xcos2k_y\\
h_{33/44}(\bm{k})&=&\epsilon_3+2t^x_{33}cosk_{x/y}+2t^y_{33}cosk_{y/x}+4t^{xy}_{33}cosk_xcosk_y+2t^{xx}_{33}cos2k_{x/y}+2t^{yy}_{33}cos2k_{y/x}\nonumber\\
&&+4t^{xxy}_{33}cos2k_{x/y}cosk_{y/x}+4t^{xyy}_{33}cosk_{x/y}cos2k_{y/x}\\
h_{55}(\bm{k})&=&\epsilon_5+2t^x_{55}(cosk_x+cosk_y)+4t^{xy}_{55}cosk_xcosk_y+2t^{xx}_{55}(cos2k_x+cos2k_y)+4t^{xxy}_{55}(cos2k_xcosk_y+cosk_xcos2k_y)\nonumber\\
&&+4t^{xxyy}_{55}cos2k_xcos2k_y\\
h_{12}(\bm{k})&=&-4t^{xxy}_{12}(sin2k_xsink_y-sink_xsin2k_y),\\
h_{13/14}(\bm{k})&=& \pm 2it^x_{13}sink_{x/y} \pm 4it^{xy}_{13}sink_{x/y}cosk_{y/x},\\
h_{15}(\bm{k})&=&2t^x_{15}(cosk_x-cosk_y)+2t^{xx}_{15}(cos2k_x-cos2k_y)+4t^{xxy}_{15}(cos2k_xcosk_y-cosk_xcos2k_y),\\
h_{23/24}(\bm{k})&=& 2it^y_{23}sink_{y/x}+4it^{xy}_{23}cosk_{x/y}sink_{y/x}+4it^{xxy}_{23}cos2k_{x/y}sink_{y/x},\\
h_{25}(\bm{k})&=& -4t^{25}_{xy}sink_xsink_y-4t^{25}_{xxy}(sin2k_xsink_y-sink_xsin2k_y),\\
h_{34}(\bm{k})&=& -4t^{34}_{xy}sink_xsink_y,\\
h_{35/45}(\bm{k})&=& 2it^x_{35}sink_{x/y}+4it^{xy}_{35}sink_{x/y}cosk_{y/x}+2it^{xx}_{35}sin2k_{x/y}.
\end{eqnarray}
The corresponding tight binding parameters are specified in unit of $eV$ as,
\begin{eqnarray}
&&  \epsilon_1=2.824439,\quad \epsilon_2=1.602385,\quad \epsilon_3=1.794899, \quad \epsilon_5=1.503426,\\
&& t^x_{11}=-0.336882, \quad t^{xy}_{11}=0.088921,\quad t^{xx}_{11}=-0.032852,\quad t^{xxy}_{11}=0.0,\quad t^{xxyy}_{11}=0.004716,\\
&& t^x_{22}=-0.161413, \quad t^{xy}_{22}=-0.027797,\quad t^{xx}_{22}=-0.007141,\quad t^{xxy}_{22}=-0.007962,\quad t^{xxyy}_{22}=-0.004822,\nonumber\\
&& t^x_{33}=-0.107864, \quad t^{y}_{33}=-0.014042,\quad t^{xy}_{33}=0.005590,\quad t^{xx}_{33}=0.022678,\quad t^{yy}_{33}=t^{xxy}_{33}=t^{xxyy}_{33}=0,\nonumber\\
&& t^x_{55}=0.038130, \quad t^{xy}_{55}=-0.018292,\quad t^{xx}_{55}=t^{yy}_{55}=t^{xxy}_{55}=t^{xxyy}_{55}=0,\nonumber\\
&& t^{xxy}_{12}=0.008613,\quad t^{x}_{13}=-0.231015,\quad t^{xy}_{13}=0.015815,\quad t^{x}_{15}=-0.011937,\quad t^{xx}_{15}=-0.003796\nonumber\\
&& t^{xxy}_{15}=0,\quad t^{y}_{23}=-0.009767,\quad t^{xy}_{23}=0.016265,\quad t^{xxy}_{23}=0.007045,\quad t^{xy}_{25}=0.004568\nonumber\\
&& t^{xxy}_{25}=0,\quad t^{xy}_{34}=0.016076,\quad t^{x}_{35}=-0.101877,\quad t^{xy}_{35}=0.008791,\quad t^{xx}_{35}=0.
\label{hopping}
\end{eqnarray}
  \end{widetext}
Without Sr doping, the chemical potential is $\mu= 2.2519$ eV with the occupation number being 8.2. With the above parameters, the obtained band structures (black lines) are shown in Fig.\ref{dfttb}, in good agreement with DFT calculations (Ni1 orbitally resolved bands). Near the Fermi level, the DOS is dominantly attributed to $d_{x^2-y^2}$, $d_{xy}$ and $d_{xz/yz}$ orbitals. For Nd$_{1-x}$Sr$_x$NiO$_2$ with $0.12\leq x\leq 0.25$ in experiments, the hole doping for the surface NiO$_2$ layer is estimated to be $0.08\leq \delta\leq 0.16$. For Nd$_{0.8}$Sr$_{0.2}$NiO$_2$, the corresponding chemical potential is $\mu=2.1339$ eV with $\delta=0.12$ and the corresponding Fermi surfaces are displayed in the main text.

\section{multi-orbital t-J model for surface NiO$_2$ layers}
\begin{figure}[tb]
\centerline{\includegraphics[width=0.8\columnwidth]{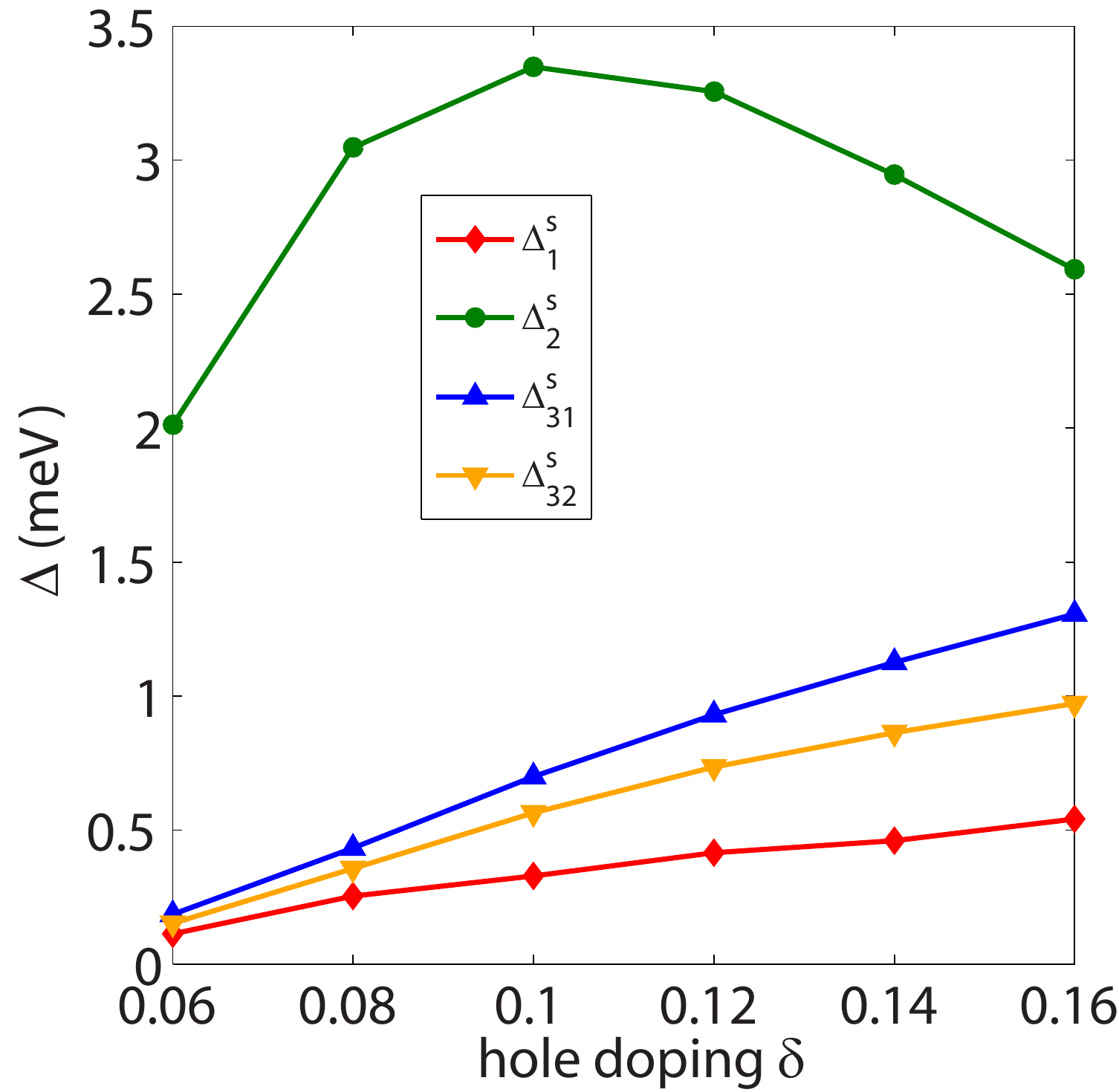}}
\caption{(color online) Orbital dependent superconducting gaps as a function of hole doping for $J^1_1=0.12$ with $J^{2,3,4}_1=J^1_1/2$.
 \label{gapdope} }
\end{figure}

In the strong-coupling limit, similar to iron based superconductors, we consider for the surface NiO$_2$ layers in-plane antiferromagnetic couplings between the Ni spins,
\begin{eqnarray}
H_{J}=\sum_{\langle ij\rangle\alpha}J^\alpha_{ij}(\mathbf{S}_{i\alpha}\mathbf{S}_{j\alpha}-\frac{1}{4}n_{i\alpha}n_{j\alpha})
\end{eqnarray}
where
$\bm{S}_{i\alpha}=\frac{1}{2}c_{i\alpha\sigma}^{\dagger}\bm{\sigma}_{\sigma\sigma'}c_{i\alpha\sigma'}$
is the local spin operator and $n_{i\alpha}$ is the local density
operator for Ni $\alpha$ orbital ($\alpha=x^2-y^2,xy,xz,yz,z^2$). $\langle ij\rangle$ denotes the in-plane nearest neighbors and the in-plane coupling is $J^\alpha_x=J^\alpha_y=J^\alpha_1$. By performing the Fourier transformation, $H_{J}$ in momentum space reads
\begin{eqnarray}
H_{J}=\sum_{\alpha\mathbf{k,k}'}V^\alpha_{\mathbf{k},\mathbf{k}'}c_{\bm{k}\alpha\uparrow}^{\dagger}c_{-\bm{k}\alpha\downarrow}^{\dagger}c_{-\bm{k}'\alpha\downarrow}
c_{\bm{k}'\alpha\uparrow},
\end{eqnarray}
with
$V^\alpha_{\mathbf{k},\mathbf{k}'}=-\frac{2J^\alpha_1}{N}\sum_{\pm}(cosk_x\pm cosk_y)(cosk'_x\pm cosk'_y)$. Here we investigate the pairing state for doped system and neglect the
no-double-occupancy constraint on this $t-J$ model and perform a mean-field decoupling. With this, the total Hamiltonian can be written as,
\begin{eqnarray} H_{MF}&=&\sum_{\mathbf{k}}\Psi_{\mathbf{k}}^{\dagger}A(\mathbf{k})\Psi_{\mathbf{k}}+\sum_{\alpha;\nu=s,d}\frac{N}{2J^{\alpha}_1}|\Delta^{\nu}_\alpha|^2,\\
A(\mathbf{k})&=&\left(\begin{array}{cc}
h(\mathbf{k}) & \Delta_{\uparrow\downarrow}(\mathbf{k}) \\
 \Delta^{\dagger}_{\uparrow\downarrow}(\mathbf{k}) & -h^{*}(-\mathbf{k}) \\
 \end{array}\right), \nonumber\\
 \Delta_{\uparrow\downarrow}(\mathbf{k})&=&\text{diag}\{\Delta_{1}(\mathbf{k}), \Delta_{2}(\mathbf{k}), \Delta_{3}(\mathbf{k}),\Delta_{4}(\mathbf{k}),\Delta_{5}(\mathbf{k}) \},\nonumber\\
\end{eqnarray}
where $\Psi_{\mathbf{k}}^{\dagger}=(\psi^\dag_{\bm{k}\uparrow},\psi^T_{-\bm{k}\downarrow})$, $\Delta_{\alpha}(\mathbf{k})  =  \Delta^s_{\alpha}(cosk_x+cosk_y)+\Delta^{d}_{\alpha}(cosk_x-cosk_y)$, and
\begin{eqnarray}
\Delta_\alpha^{s/d} &=& -\frac{2J^\alpha_1}{N}\sum_{\mathbf{k}'}d^\alpha_{\mathbf{k}'\uparrow}(cosk'_x\pm cosk'_y),
\end{eqnarray}
with $d^\alpha_{\mathbf{k}'\uparrow}=\left\langle
c_{-\mathbf{k}'\alpha\downarrow}c_{\mathbf{k}'\alpha\uparrow}\right\rangle
$. $A(\mathbf{k})$ can be diagonalized by a unitary transformation
$U_{\mathbf{k}}$ with and the Bogoliubov quasiparticle eigenvalues $E_{m+5}=-E_{m}$ with $m=1,2,3,4,5$. The self-consistent gap equations are
\begin{eqnarray}
\Delta^{s/d}_{\alpha} & = & -\frac{2J^\alpha_1}{N}\sum_{\mathbf{k},m}(cosk_x\pm cosk_y)U_{\alpha+5,m}^{*}(\mathbf{k})\nonumber\\
&& \times U_{\alpha,m}(\mathbf{k})F[E_{m}(\mathbf{k})],
\end{eqnarray}
where $F[E]$ is Fermi-Dirac distribution function, $F[E]=1/(1+e^{E/k_{B}T})$. We further introduce $\Delta^{s/d}_{31}=\frac{1}{2}(\Delta^{s}_{3}\pm \Delta^{s}_{4}+\Delta^{d}_{3}\mp \Delta^{d}_{4}) $ and $\Delta^{s/d}_{32}=\frac{1}{2}(\Delta^{s}_{3}\pm \Delta^{s}_{4}-\Delta^{d}_{3}\pm \Delta^{d}_{4}) $ for $d_{xz/yz}$ orbitals. The above equations can be solved self-consistently, varying the doping and the value of $J^\alpha_1$. In the calculations, we used  $J^{2,3,4}_1=J^1_1/2$ and omitted the exchange coupling for $d_{z^2}$ orbital i.e. $J^5_1=0$. The orbital dependent superconducting gaps as a function of hole doping is displayed in Fig.\ref{gapdope}, where the extended $s$-wave pairing is always dominant and $d_{xy}$ has the largest gap size.

With including interlayer couplings between the surface and bulk NiO$_2$ layers in NiO$_2$ terminated surface, we perform calculations to study pairing states. The electronic structure of bulk NiO$_2$ layers can be described by one-band model and we consider the Hamiltonian $H=H_0+H_1+H_t$, where $H_1$ describes the bulk layer with a hole pocket around $\bar{\text{M}}$ and $H_t$ is the tunneling between the two layers. In the calculations, the band structure of bulk NiO$_2$ layer is approximated by that at $k_z=0$ plane and the corresponding parameters can be found in Ref.\onlinecite{WuPRB2020}. For interlayer couplings, we include the nearest-neighbor (NN) coupling between $d_{x^2-y^2}$ orbitals in two layers $t^z_{16}=-0.0368$ eV and the fourth NN coupling between $d_{xy}$ and $d_{x^2-y^2}$ orbitals in two layers $t^{xxy}_{26}=0.015$ eV (the NN to third NN hoppings vanish). By including the nearest neighbor exchange interactions for both layers, we solve the system within a mean-field decoupling similar to the case of surface layers and find the $s_{\pm}-id_{x^2-y^2}$ pairing state is energetically more favorable. We have further included two or three bulk NiO$_2$ layers in our calculations and find that $s_{\pm}-id_{x^2-y^2}$ is always favored.

\end{document}